\documentstyle[12pt,aasms4]{article}

\begin{document}

\title{Balmer and He I absorption in the nuclear spectrum of NGC 4151
\footnote{Based on observations made with the NASA/ESA Hubble Space Telescope.
STScI is operated by AURA Inc, under NASA contract NAS5-26555.}}

\author{J. B. Hutchings\footnote{Guest user, Canadian Astronomy
Data Centre, which is operated by the HIA, NRC of Canada}}

\affil{Herzberg Institute of Astrophysics, National Research Council 
of Canada, 5071 W. Saanich Rd, Victoria, B.C., V9E 2E7, Canada;
john.hutchings@nrc.ca}

\author{D.M. Crenshaw}
\affil{Dept of physics and astronomy, Georgia State University,
Atlanta, GA 30303}

\author{S.B. Kraemer, J.R. Gabel}
\affil{The Catholic University of America/IACS, NASA/Goddard Space Flight
Center, Code 681, Greenbelt, MD, 20771}

\author{M.E. Kaiser}
\affil{Dept of physics and astronomy, Johns Hopkins University,
3400 North Charles St, Baltimore, MD 21218}

\author{D. Weistrop}
\affil{Dept of Physics, University of Nevada, Las Vegas, NV89154-4002}

\author{T.R. Gull}
\affil{Nasa Goddard Space Flight Center, code 681, Greenbelt, MD 20771-5302} 
\begin{abstract}

   Spectra taken with the Space Telescope Imaging Spectrograph (STIS)
allow accurate location and extraction of
the nuclear spectrum of NGC 4151, with minimal contamination by extended
line emission and circumnuclear starlight. Spectra since 1997 show that 
the P Cygni Balmer and He I absorption seen previously in low nuclear states, 
is present in higher states, with outflow velocity that changes with 
the nuclear flux. The phenomenon is discussed in terms of some of the 
absorbers seen in the UV resonance lines, and
outflows from the central source and surrounding torus.

\end{abstract}

\keywords{galaxies: Seyfert -- galaxies: individual (NGC 4151)}

\section{Introduction}

   NGC 4151 is the brightest Seyfert 1 type galaxy, and has been studied
in considerable detail at all wavelengths. In reference to the nuclear
region, the Hubble Space Telescope has been instrumental in resolving the
innermost few arcseconds, and revealing the spatial and velocity structure 
of the narrow emission line gas (see e.g. Hutchings et al 1998, Kaiser et al
2000, Nelson et al 2000, Crenshaw et al 2000). The detailed picture that 
emerged, for NGC 4151 and other Seyferts, is
of a hollow biconical outflow of narrow-line clouds. In the case of
NGC 4151, the high velocity radio jets lie along the cone axis, and 
our line of sight lies close to the edge of the
approaching cone. This scenario was put forward earlier by, for example, 
Pedlar et al (1993) and Boksenberg et al (1995).

   NGC 4151 is also known to show flux variations over a factor of ten or more,
and has been the subject of echo-mapping observational campaigns. These have
shown the inner broad emission line region to have an extent of several light
days (see overview by Peterson et al 1998). The brightness and spatial 
extent of the narrow
emission lines has made it difficult to isolate the nuclear spectrum and
emission lines. It was noted originally by Anderson and Kraft (1969)
that there are shortward shifted absorptions in H$\gamma$ and the metastable
He I $\lambda$3888 line. Anderson (1974) followed up with further data and
a discussion that suggested a connection between the continuum flux and
the absorption strength. This has been poorly documented since, but Sergeev,
Pronik, and Sergeeva (2001) give a summary of observations over 11 years
that show the absorptions are present in a nuclear low state in 1999.
No systematic study has been made of the absorptions, perhaps because
ground-based observing conditions cause a large range of contamination 
by the circumnuclear flux, both line and continuum. 

   The long slit (or slitless) spectroscopic capability of STIS, along with
the spatial resolution of the HST, has made it possible to obtain and study 
the nuclear spectrum consistently and cleanly. There is extended narrow
line emission with many velocity components, even within the central arcsec,
which can affect the overall line profiles, if included.
In this paper we discuss the series
of visible range nuclear spectra from STIS, that fortuitously cover a wide
range of nuclear flux variations. We are particularly interested in the outflow
absorption that is seen in the strong Balmer and the metastable He I line. 

   Outflow is also seen in higher velocity emission line clouds near the
nucleus (Hutchings et al 1999), multiple shifted absorption lines in 
C IV and other UV resonance lines (Weymann et al 1997, Crenshaw et al 2000,
Kriss et al 2002), and warm absorbers seen in X-ray data (see e.g. Schurch 
and Warwick 2002). A full picture 
of the different outflows has yet to emerge, and this paper adds further
information to the inventory. 

\section{Observations and data}

   Table 1 shows the observations used in this paper, along
with some principal measures. With one exception (June 2000), the 
observations were executed as programs by the authors, so the data 
are known to be suitable for this investigation. In addition to the spectra
listed in table 1, we inspected and measured associated STIS spectra 
covering the far-UV to 1 micron, to obtain a complete picture of the nuclear
spectrum at the same times. The spectral resolutions at H$\beta$ are
$\sim$800 for the G430L and $\sim$8000 for G430M spectra. 

The nuclear flux varied considerably over the
timespan covered, including some unusually low states. The nuclear spectrum 
was extracted from the long slit (or slitless) data by using the detector rows
that covered the continuum, clearly detected. These were collapsed to a single
spectrum, and used for further measurements. The data were
retrieved from the CADC with on-the-fly calibration, and also extracted
using CALSTIS from the STIS team database. Two observations
were made nominally offset by 0.09" from the nucleus (June 1998 and June 1999).
The nuclear continuum is clearly present in the spectra, but the slit
should have lost some of the flux. From the nuclear cross-sections
in the other spectra, we
estimate that the continuum fluxes for these spectra may be underestimated by
a factor 2.5, and this factor is included in the Table 1 values. We note that
the overall correlation with nuclear flux is not altered by this
correction, or by an uncertainty of a further factor two in either direction.  

    Figure 1 shows the average of the five G430L spectra as extracted this
way, after each had been normalised to the continuum, and Figure 2 shows 
the comparison of the two G430M spectra (not normalised, as the coverage
was not sufficient to establish the continuum level). 

   The asymmetry of the H$\beta$ profiles (and other Balmer lines) is apparent
in Figures 1 and 2.  The profiles consist of a broad emission, a narrow
emission, and a shortward absorption trough. The H$\gamma$ line is blended
with [O III] 4363\AA~ emission, and H$\delta$ is blended with [S II] emission.
We also note the absorption feature near 3900\AA, which is unique in the 
nuclear spectrum.

   In Figure 3, the Balmer lines H$\beta$, H$\gamma$, and H$\delta$ are
superposed in velocity space, after scaling to the same broad emission
line profile peaks. The Figure also sketches in a symmetrical broad
emission profile for the profiles from each observation. This was derived
by folding the profile about velocities near to zero and matching the 
unblended parts of them on each side. The agreement among the three Balmer
lines is notable and lends confidence in the result. Note that in all cases
there is an apparent shortward absorption, and that it is much more obvious
in the later spectra, when the continuum level was very low. In the June 1999
spectrum, the Balmer absorption extends to higher velocities, beyond the
deep minimum that corresponds with the main feature seen in other low-state
spectra. We measured this absorption as two separate features in this spectrum.

We also note that the centres of the symmetrical broad Balmer emission 
profiles are consistently longward of the 1000 km s$^{-1}$ generally 
quoted for NGC 4151, by an average of 350 km s$^{-1}$. (This standard 
value is presumably derived from the mean of many blended narrow 
emission peaks of different velocity over the nuclear region
and hence somewhat arbitrary.) It may thus be that the velocities
recorded in Table 1 should be more negative by this amount, to represent
velocities with respect to the central BLR. However, we note that in earlier
bright epochs, the broad emission line profiles are stronger on the shortward
side of the nominal redshift (see Sergeev et al 2001), so that there may be 
changes in the broad profiles
(which may be caused by changing obscuration of the redshifted outflowing
matter on the other side of the nucleus), and our symmetrical
assumptions in Figure 3 may not apply to other epochs.

   Table 1 shows the measures of the Balmer absorption, as well as the
equivalent width of the He I absorption, and measures of the
He II 4686\AA~ peak. This latter has a broad
emission component too, but the peak (as all other narrow emissions) largely
varies in EW as a result of the continuum changes, and the values serve as
a consistency check on the continuum flux numbers. We define the V$_{min}$
value as the turning point of a parabola fit to the absorption profile, and
V$_{edge}$ as the shortward limit of absorption as illustrated in Figure 3.

   Finally, we measured the velocity of the minimum of the $\sim$3900\AA~ 
absorption, assuming the identification is He I 3888\AA. The decrement of the
Balmer absorption and emission suggest this identification, plus the facts
that this metastable state line is known to arise in high density outflows, 
and that the
velocities do not agree with the other Balmer absorptions. Anderson and 
Kraft (1969) made the same argument. The absorption
lies between two emission lines (see Fig 1) so a concern is that changes
may be distorted by blending. However, we find no changes in the absorption
or emission FWHM in the sense that the velocity changes would require if the
lines are blended, so conclude that the absorption feature is resolved 
with the G430L spectra. We measured the absorption EW with respect to the
continuum beyond the neighbouring  emissions, which may
underestimate the true value. However, the measurements are well defined and
consistent, and also  show a similar variation to the H$\beta$ absorption. 

If we use the [Ne III] or [O III] emission as a wavelength fiducial 
instead of the data calibration, we find the absorption velocities may 
change by up to 50 km s$^{-1}$. This will not reduce the significance
of the H$\beta$ changes, but are comparable with the He I range in Table 1. 
Variable inclusion of different narrow emission components in the different
spectra are a more likely explanation of their velocity scatter, however. 
In Figure 4 we show the line velocities and absorption strengths as a 
function of the continuum flux at 4800\AA. We discuss the correlations
further below. 

  In addition to the spectra listed in Table 1, we extracted spectra from
the G750L, G230M, G140L, and G140M gratings, taken at the same times.
These spectra show that the derived fluxes show no discontinuities from
1200\AA~ to 1 micron. We also inspected and measured major line features
in these wavelengths, including H$\alpha$, Mg II 2800\AA, and C IV, Si IV, N V,
and Ly$\alpha$ in the far UV. The H$\alpha$ line is strong and blended
with [N II], so is not useful for studying the absorption (although it
clearly is present in the form of asymmetry of the shortward side of
the peak). The far UV lines have been discussed in
detail by several other authors, and we discuss below possible 
correlation with the
varying Balmer and He I absorption in the G430 spectra. The C IV profile
shows many absorption components shortward of line centre, and these appear
as a single smooth profile in the low dispersion spectra. However, our
G430M spectra show clearly that the H$\beta$ absorption is a single broad
feature and not resolvable into sharp components as seen in the UV
resonance lines. This too relates to our discussion below.

   We measured the absorption FWHM values where possible - i.e. the He I
absorption, and the deep absorption profiles in the 1999 and 2000 spectra.
The He I line is consistent with a value of 460 km s$^{-1}$ for all cases,
while the Balmer absorption is 340 km s$^{-1}$ in the G430L spectra and
420 km s$^{-1}$ in the July 2000 G430M spectrum. Smoothing the G430M
spectrum to the resolution of the G430L does not alter the FWHM value, so
the profiles are resolved in all spectra.

\section{Discussion}

  The presence of Balmer and He I $\lambda$3888 absorption indicates
the presence of relatively high density and low ionisation outflowing material.
Furthermore, the outflow is apparently connected to the variations in the
continuum flux.  While the H$\beta$ absorption has been noted before in
low continuum states, it has not been isolated well from extended line
emission, or correlated with the nuclear variations.

  We find that there is an asymmetry in the Balmer profiles at all nuclear
flux states, that may be measured as an outflow (P Cygni) absorption, and
that in fact the absorbed flux is largest when nucleus is in a high state.
We also find that the velocity of the outflow is highest in the high nuclear
state. The He I absorption shows a similar correlation but at lower
outflow velocities.  While the Balmer line measures in the
high nuclear states depend on assuming the broad profile is symmetrical
(Figure 3), and as the broad profiles do on other occasions have
considerable blue-ward asymmetry, we may be wary of these measured values.
However, the correlation with nuclear flux, the close agreement among 3 Balmer
lines, and the changes in the (broad-component-free) He I line, suggest 
the effects are real.

  There are outflow absorptions seen in the far-UV resonance lines, that
have been discussed in detail (e.g. Weymann et al 1997, Crenshaw et al 2000,
Kraemer et al 2001). Most of these absorbers are narrow and do not change 
by much, if at all, and may arise in the clouds similar to those responsible 
for the narrow emission lines, from an extended region outside the BLR. 
However, there are some broader absorbers in the UV lines - in particular
component D+E in Kraemer et al (2001), which has velocity -490 km s$^{-1}$ 
and FWHM 435 km s$^{-1}$. This component has a high density of absorbing
material and may give rise to Balmer absorption too. The H$\beta$ EW of
3.2 (July 2000) implies a column of 1.5 x 10$^{14}$cm$^{-2}$, while the EW for
He I of 1.4 (May 2000) implies 1.8 x 10$^{14}$cm$^{-2}$ if they are
associated with the UV component D+E responding to changes in the ionising
continuum. It seems likely that this UV absorber is the same as
that causing the low velocity strong Balmer absorption in the nuclear low 
state. However, the connection with the higher velocity Balmer absorption,
and the lower velocity He I absorption is not clear.

  In the higher nuclear states, the Balmer and He I velocities increase,
as seen in Figure 4. These Balmer profiles do not appear to be composed
of two or more components, and the complex absorption spectra in the
UV resonance lines do not appear to include such changes. On the other
hand the He I absorber shows much less change in velocity.
Thus, association of the Balmer and He I absorbers with
components of the highly ionised species of C IV, Si IV, N V is unclear. 
The FUSE spectrum of NGC 4151 in a low flux state shows
smooth broad absorption profiles in O VI. These will be discussed in detail
by Kriss et al (2002), but for this discussion, we assume they arise in an
accelerating flow from the central disk, as discussed for NGC 3516 by
Hutchings et al (2001). 

   NGC 4151 is often noted as being a marginal Sy 1 type and the outflow models
for the NLR gas suggest that the line of sight lies close to the edge of the
opening cone of the ionising radiation (see e.g. Crenshaw et al 2001). 
This means that whatever is defining
the cone lies close to the line of sight. This is generically referred to as
the obscuring torus. It is very reasonable to propose that the nuclear
activity is causing some erosion of the edge of the torus, and this 
may be where the outflows we see in H and He I arise. The radiation
effects that drive the flow will vary with the nuclear flux, and lead to
the velocity/flux correlation we see in Figure 4, with some time delay. 
A sudden drop in nuclear flux may leave a weakening broad absorption 
profile that lasts until the high velocity flow has dispersed: this is
possibly what we see in the June 1999 profiles, since UV spectra from 
a few months earlier show the nuclear flux to be much higher. By contrast,
the low state of June 2000, was preceded by low flux in April 2000.

   The lower outflow velocity seen in the He I absorption must be significant.
It suggests an acceleration in a cooling medium, as in stellar winds. It is
interesting that Anderson and Kraft (1969) saw the same velocity difference,
reinforcing our
conclusion that the He I absorption is not significantly blended with the
neighbouring [Ne III] emission line. Anderson (1974) reports structure within 
the He I absorption, which is comparable to the noise in his spectra (and
may also involve variable off-nuclear contamination). This is not 
resolvable in our G430L spectra. However, we note that the ten times higher
resolution G430M spectra of H$\beta$ show very smooth absorption profiles
(see Fig 2).

   The flow velocities are similar to those seen in the NLR and the
associated sharp absorptions in the UV resonance lines. They are smaller than
the higher velocity flows seen in the inner NLR (Hutchings et al 1999),
even without the projection effects that must apply to emission line clouds.
They are also somewhat smaller than the flow velocities seen in massive
star winds (e.g. Fullerton et al 2000). The flux from the nucleus of
NGC 4151 is in the range of 100 to 10$^5$ times that of an OB star, so the
radiation pressure would be similar at a distance of 3.5x10$^{15}$ cm, or
0.001 pc, if we want the same process to apply. The time lag over this 
distance is of order one day, which places it within the BLR by the echo 
mapping results. However, we see no unambiguous evidence that the higher
velocity flow shows 
up in the UV resonance lines, which are the principal drivers of stellar
winds. Thus, it is possible that the flow reported here may arise in a
more distant location, and that some other force than central radiation 
may drive it.

 It thus seems possible that the outflow
arises in a high density region (such as the torus edge) which is heated
by both the nuclear radiation and outflowing material. 
The mild heating and acceleration we see may arise
by entrainment in the biconical outflow. In very high nuclear states (not
sampled in the data in this paper), the broad emission line profile changes
asymmetrically, and perhaps the low ionisation outflow velocity is higher,
so that it may be more difficult to identify it in the line profiles. It will
be instructive to continue to monitor the nuclear visible spectrum through
higher nuclear flux states. The special NGC 4151 line of sight geometry 
may be a valuable clue on the origins of the nuclear outflows.

\clearpage
\centerline{Figure captions}

1. Average low dispersion (G430L) nuclear spectrum of NGC 4151 from data 
January 1998 to June 2000, normalized to the continuum defined by line-free
regions. The principal emission lines, and He I 3888\AA, are identified,
redshifted by 1000 km s$^{-1}$. The spectra exclude all light beyond
$\sim$0.15 arcsec from the nucleus.

2. Comparison of the nuclear spectra from the G430M grating, from high and low
nuclear flux states. Note that the H$\beta$ absorption is a single
broad feature and not composed of several narrow absorptions, as seen in 
the UV resonance lines.

3. Normalised Balmer profiles in velocity space, flux-scaled to match their
broad line components. The narrow line components are relatively stronger
in the later spectra, when the nuclear flux was low. In these, the [O III] 
$\lambda$4959 line is truncated for easier viewing. Similarly, the [O III] 
$\lambda$4363 and [Si II] emissions, near H$\gamma$ and H$\delta$
respectively, are truncated.
A symmetrical broad emission profile has been sketched in to outline
the Balmer absorption. This absorption is seen to agree well between the
three Balmer lines. The average centre of the symmetrical broad profiles  is marked, which is somewhat longward of the nominal NGC 4151 redshift
of 1000 km s$^{-1}$. The numbers in parentheses next to the dates give the
nuclear flux at 4800\AA~ in units of 
10$^{-14}$ erg \AA$^{-1}$ cm$^{-2}$ sec$^{-1}$.

4. Absorption line measures from Table 1 plotted against continuum flux at
4800\AA. Note the trend to higher outflow velocity with increasing nuclear 
flux. While the absorption equivalent widths drop, the absorbed flux
rises with increasing nuclear flux. The circled points refer to the high
velocity wing of the absorption in the June 1999 spectrum. The lines are
fits to the data as labelled, except for H$\beta$ absorption, where the
points are joined.

\clearpage
\centerline{References}

Anderson K.S., and Kraft R.P., 1969, ApJ, 158, 859

Anderson K.S., 1974, ApJ, 189, 195

Boksenberg A., et al, 1995, ApJ, 440, 151

Crenshaw D.M. et al, 2000, AJ, 120, 1731

Fullerton A.W. et al, 2000, ApJ, 538, L43

Hutchings J.B. et al 1998, ApJ, 492, L115

Hutchings J.B. et al 1999, AJ, 118, 2101

Hutchings J.B. et al 2001, ApJ, 559, 173

Kaiser M.E. et al, 2000, ApJ, 528, 260

Kraemer S.B. et al, 2001, ApJ, 551, 671

Kriss G.A. et al 2002 (in preparation)

Nelson C.H. et al, 2000, ApJ, 531, 257

Peterson B.M. et al, 1998, PASP, 110, 660

Pedlar A., et al, 1993, MNRAS, 263, 471

Sergev S.G., Pronik V.I., and Sergeeva E.A., 2001, ApJ, 554, 245

Schurch N.J. and Warwick R.S., 2002, astro-ph/0204057 (MNRAS in press)

Weymann R.J., Morris S.L., Gray M.E., Hutchings J.B., 1997, ApJ, 483, 717

\begin{deluxetable}{rcccccccccc}
\footnotesize
\tablecaption{NGC 4151 nuclear spectra}
\tablehead{\colhead{Date} &\colhead{JD} &\colhead{G430,Slit}
&\colhead{flux\tablenotemark{a}} 
&\multicolumn{4}{c}{H$\beta$ absorption measurements}
&\multicolumn{2}{c}{He I ~$\lambda$3888} &\colhead{$\lambda$4686}\\
&\colhead{245..} &\colhead{~~~~"} &\colhead{4800\AA} 
&\colhead{EW} &\colhead{f$_{abs}$\tablenotemark{b}}
&\colhead{V$_{edge}$} &\colhead{V$_{min}$}  &\colhead{V$_{min}$}
&\colhead{EW} &\colhead{EW}\\ 
&&&&\colhead{(\AA)} 
&&\multicolumn{3}{c}{Radial vel.\tablenotemark{c} ~(km s$^{-1}$)} 
&\colhead{(\AA)} &\colhead{(\AA)}}

\startdata

15 Jul 1997 &0645&M, 50  &6.e-14 &0.6 &1.3e-13 &-1850 &-1200 &-    
&- &1.1\cr
8  Jan 1998 &0822&L, 0.1 &7.e-14 &0.4 &1.4e-13 &-2000 &-1200 &-530 
&0.6 &1.1\cr
10 Feb 1998 &0855&L, 0.1 &6.e-14 &0.6 &1.4e-13 &-2200 &-1100 &-580 
&0.8 &1.3\cr
1  Jun 1998\tablenotemark{d} &0966&L, 0.1 &2.5e-14 &0.8 &1.1e-14 &-1600 &-750  &-365 
&1.5 &1.4\cr
4  Jun 1999\tablenotemark{d} &1334&L, 0.1 &7.e-15 &2.6 &2.0e-14 &-920  &-460  &-370 
&1.8 &4.3\cr
                                   &&&&6.2 &4.3e-14 &-2250 \cr
24 May 2000 &1689&L, 0.1 &1.e-14 &1.8 &4.0e-14 &-1135 &-605  &-495 
&1.4 &5.0\cr
2  Jul 2000 &1728&M, 0.2 &6.e-15 &3.2 &7.0e-14 &-990  &-500  &-
&-     &-\cr

\enddata

\tablenotetext{a}{Flux in erg \AA$^{-1}$ cm$^{-2}$ sec$^{-1}$}
\tablenotetext{b}{Absorbed flux in erg cm$^{-2}$ sec$^{-1}$}
\tablenotetext{c}{RV wrt redshift of 1000km s$^{-1}$}
\tablenotetext{d}{Flux corrected by 2.5 for slit offset}
\end{deluxetable}

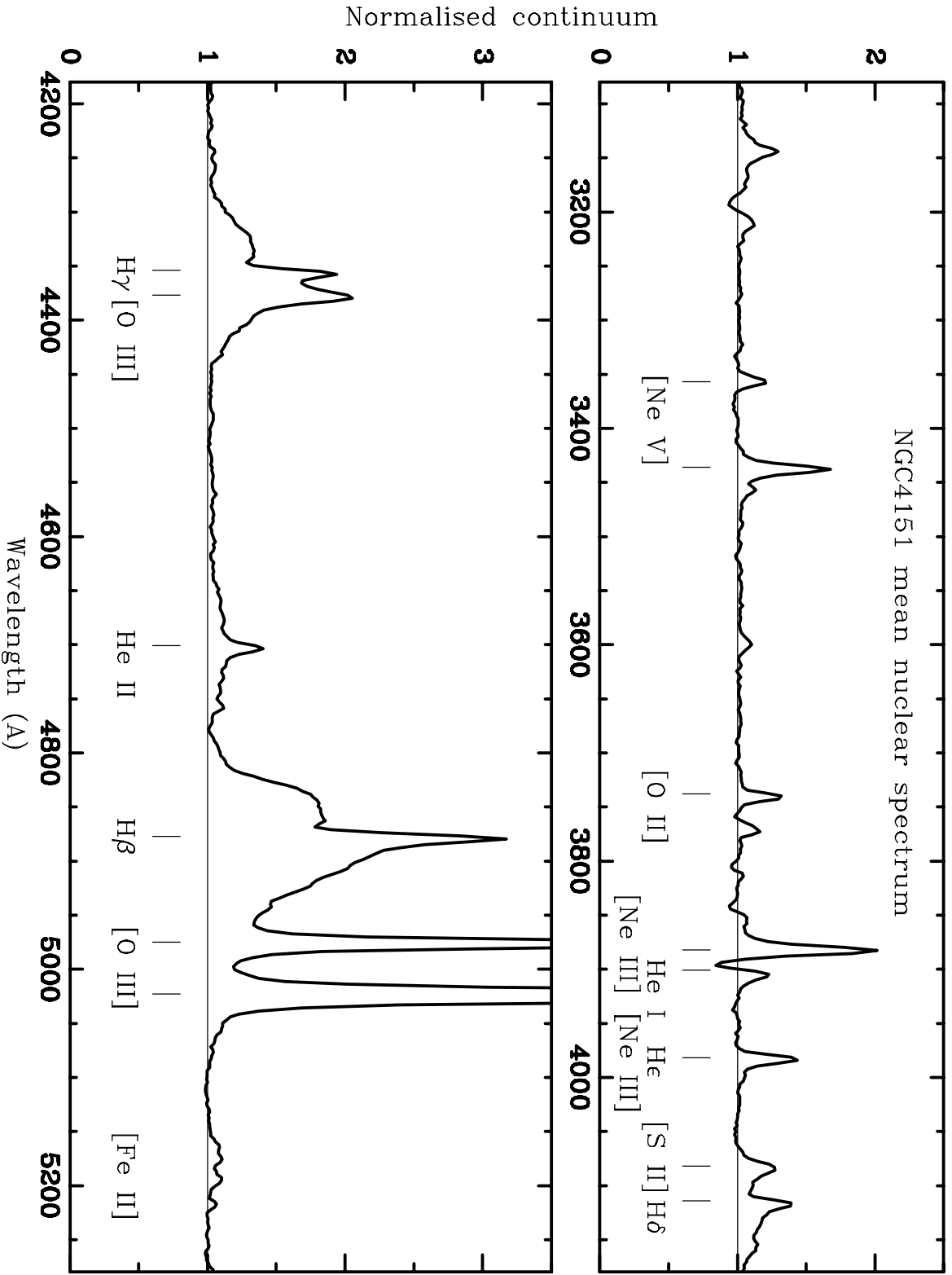
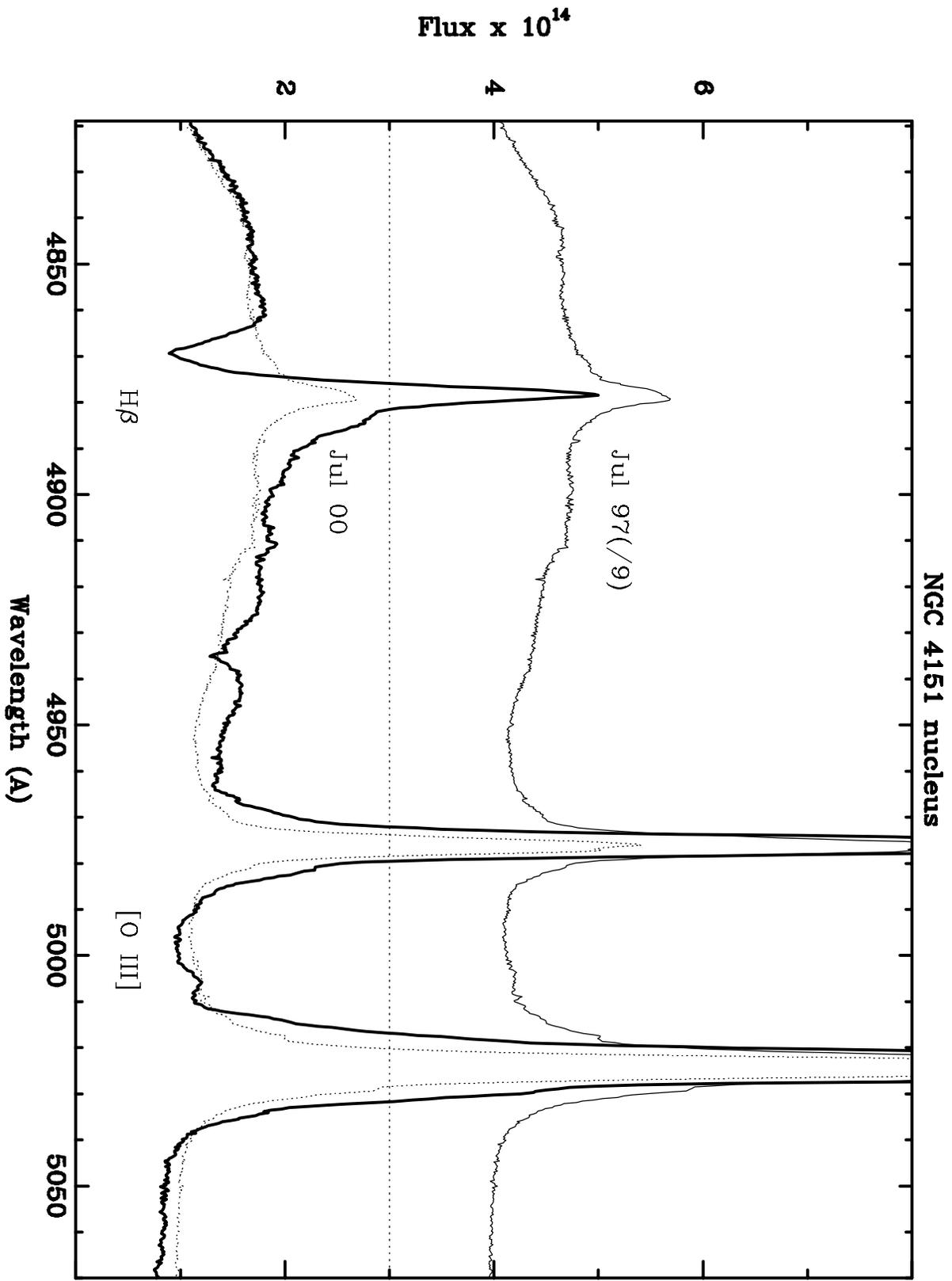
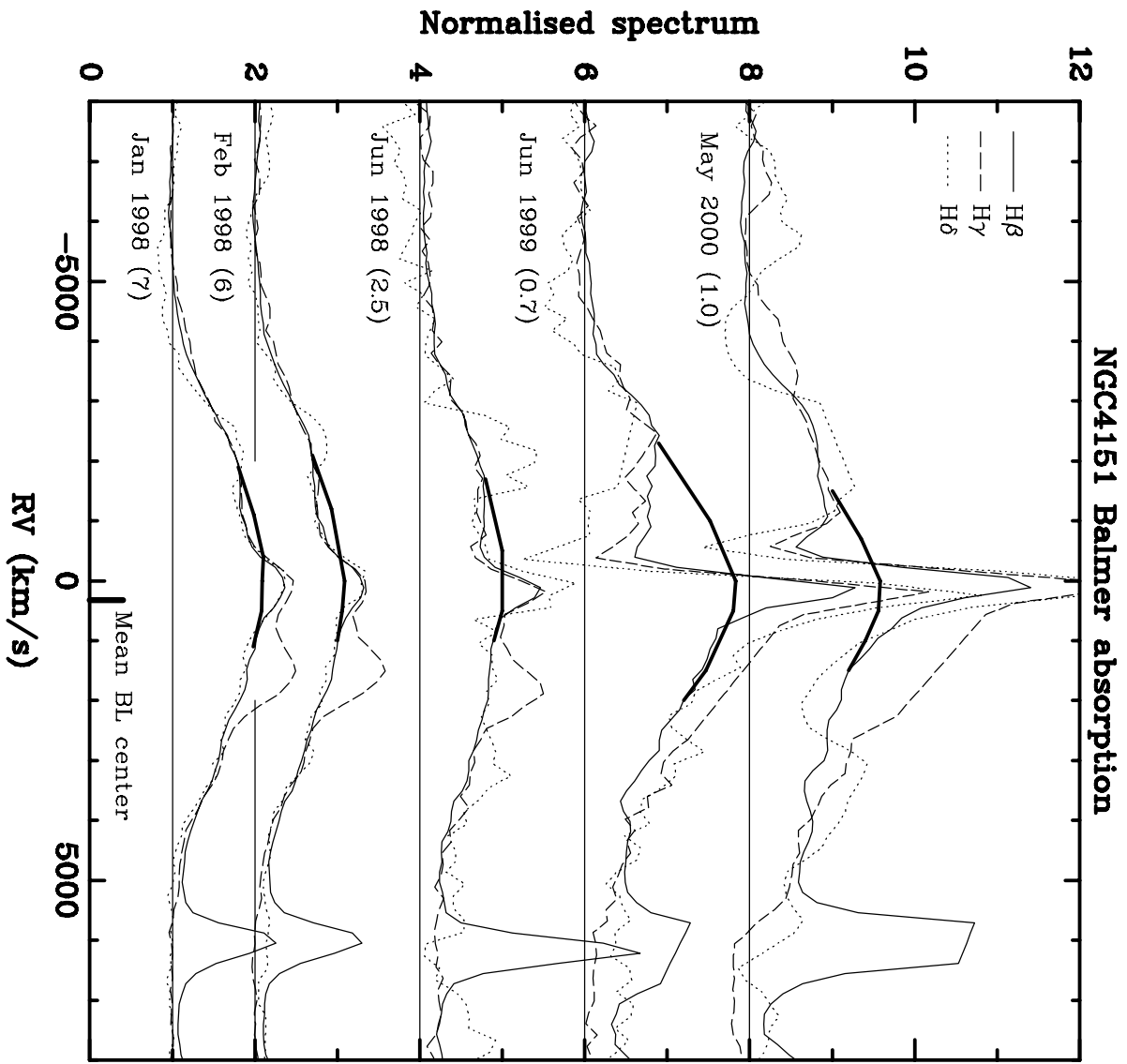
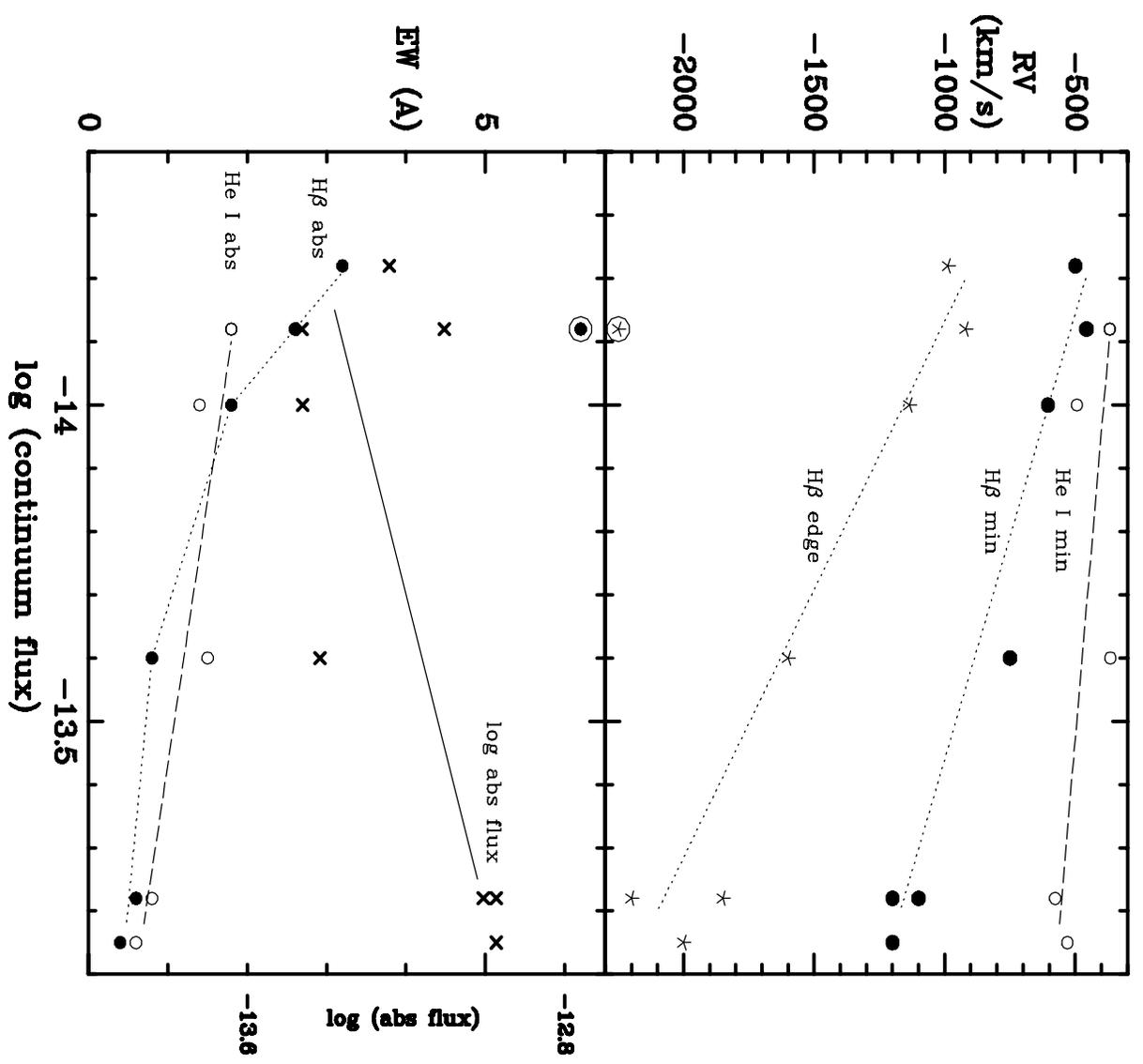
\end{document}